\documentclass[aps,prl,twocolumn,reprint,superscriptaddress,secnumarabic,amssymb]{revtex4}
\usepackage{graphicx}
\usepackage{dcolumn}
\usepackage{bm}
\usepackage{color}
\usepackage{bbold}
\usepackage[breaklinks=true,colorlinks,citecolor=blue,linkcolor=blue,urlcolor=blue]{hyperref}
\usepackage[normalem]{ulem}
\usepackage{color}
\usepackage{amsmath}
\usepackage{enumitem}

\newcommand{\bk}{{\bm k}}

\newcommand{\bq}{{\bm q}}
\newcommand{\bd}{{\bm d}}

\newcommand{\br}{{\bm r}}

\newcommand{\bR}{{\bm R}}

\newcommand{\ba}{{\bm a}}

\newcommand{\bsig}{{\bm \sigma}}
\newcommand{\btau}{{\bm \tau}}

\newcommand{\fref}[1]{Fig.\hspace{0.025in}\ref{#1}}

\newcommand{\eref}[1]{Eq.\hspace{0.025in}(\ref{#1})}

\newcommand\wordcount{
    \immediate\write18{texcount -sub=section \jobname.tex  | grep "Section" | sed -e 's/+.*//' | sed -n \thesection p > 'count.txt'}
(\input{count.txt}words)}

\begin{document}
\title{Antichiral edge states in a modified Haldane nanoribbon}

\author{E. Colom\'es}
\email{enrique.colomes@uab.es}
\affiliation{Departament d'Enginyeria Electr\`onica, Universitat Aut\`onoma de Barcelona, 08193-Bellaterra (Barcelona), Spain}
\affiliation{Department of Physics and Astronomy and Quantum Matter Institute, University of British Columbia, Vancouver, British Columbia, Canada V6T 1Z4}
\author{M. Franz}
\email{franz@phas.ubc.ca}
\affiliation{Department of Physics and Astronomy and Quantum Matter Institute, University of British Columbia, Vancouver, British Columbia, Canada V6T 1Z4}

\begin{abstract}
Topological phases of fermions in two-dimensions are often characterized by chiral edge states. By definition these propagate in the opposite directions at the two parallel edges when the sample geometry is that of a rectangular strip. We introduce here a model which exhibits what we call ``antichiral'' edge modes. These propagate {\em in the same direction} at both parallel edges of the strip and are compensated by counter-propagating modes that reside in the bulk.  General arguments and numerical simulations show that backscattering is suppressed even when strong disorder is present in the system. We propose a feasible experimental realization of a system showing such antichiral edge modes in  transition metal dichalcogenide monolayers.
\end{abstract}

\maketitle

\begin{figure}[t]
\includegraphics[width = 7.5cm]{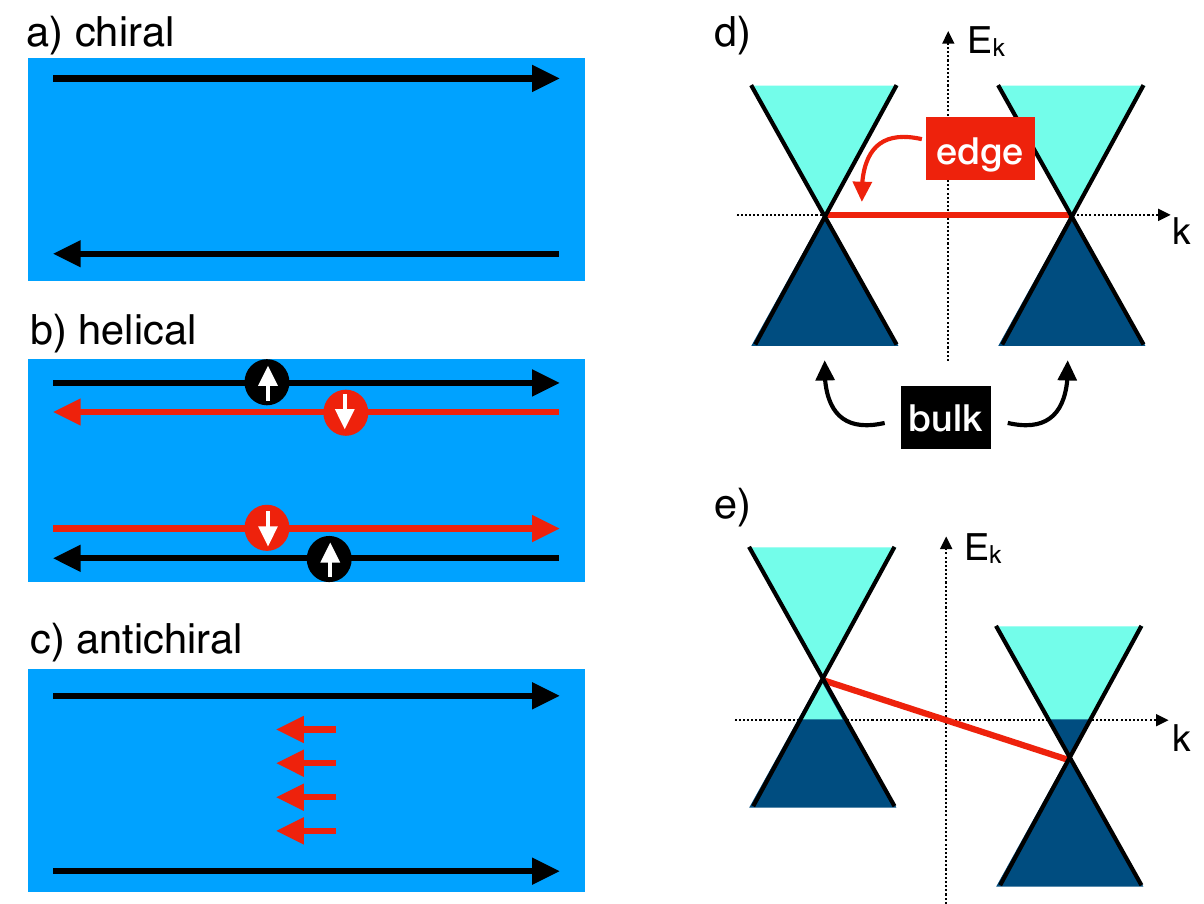}
\caption{Schematic of a topological state with (a) chiral, (b) helical and (c) antichiral edge modes. Panel (d) illustrates the low energy dispersion of a zigzag graphene nanoribbon with a dispersionless edge state (red) connecting two Dirac points. Applying the pseudoscalar potential in (e)  offsets the Dirac points in energy and causes the edge mode to disperse. 
}\label{fig1}
\end{figure}

\emph{Introduction.}-- Topologically protected edge states in two-dimensional (2D) fermionic systems occur in two common varieties.  {\em Chiral} edge modes are found in systems with broken time reversal symmetry such as the quantum Hall or Haldane insulators \cite{laughlin1,haldane}. In a strip geometry the bulk is gapped and the protected edge modes are counter-propagating as illustrated in  Fig.\ \ref{fig1}a. {\em  Helical} edge modes occur in time reversal invariant systems, such as 2D quantum spin Hall effect (QSHE) \cite{kane, hasan_rev} and can be regarded as two superimposed copies of Haldane insulators related by time reversal symmetry. In this case bulk is also gapped and each edge contains a pair of counter-propagating spin filtered states illustrated in  Fig.\ \ref{fig1}b. In both cases backscattering is suppressed exponentially with the width $W$ of the strip and leads to effectively dissipationless edge transport in the limit of large $W$ \cite{remark1}. Such loss-free transport has obvious technological potential and underlies much of the current interest in topological states of matter.

In this Letter we ask the following question: Is it possible to have a 2D fermionic system with co-propagating edge modes illustrated in  Fig.\ \ref{fig1}c? A simple consideration shows that such ``antichiral'' edge modes cannot exist in a system with a full bulk gap. This is because the number of left and right moving modes in any finite system defined on the lattice must be the same. Only then one can define a legitimate band structure with full Brillouin zone periodicity.  We show, however, that antichiral edge modes indeed can exist in 2D semimetals where gapless bulk states supply the required counter-propagating modes. The edge modes are still topologically protected, much like Fermi arcs in 3D Dirac and Weyl semimetals \cite{wan1,vafek1}. They also carry nearly dissipationless currents although the exponential protection against backscattering is replaced by a power law due to the extended nature of the counter-propagating bulk modes.

A similar situation has been encountered recently \cite{dless1,dima,Beenakker} in a 3D Weyl semimetal wires, where the energy dispersion is modified such that conducting surface modes propagate in one direction only and bulk modes propagate in the other.  In such a ``topological coaxial cable'' backscattering is strongly suppressed because bulk and surface modes are spatially separated. Reliable numerical simulations of this system with disorder are however quite challenging owing  to its 3D nature. Here we construct a 2D system with analogous physical properties in which numerical results can be obtained up to very large sizes.

A system we consider in this Letter takes inspiration from a graphene nanoribbon with zigzag edges \cite{rise}. It is well known that such a nanoribbon exhibits dispersionless edge states -- zero-energy flat bands -- that span projections of two inequivalent Dirac points onto the 1D Brillouin zone characterizing the ribbon \cite{fujita1,sigrist1,guinea}, Fig.\ \ref{fig1}d. These edge modes are topologically protected in a similar way as the Fermi arcs in 3D Dirac and Weyl semimetals \cite{ryu1,delplace1,kharitonov1}. The key idea is to add a pseudoscalar potential term to the Hamiltonian $H_0$ describing such a ribbon so  that the two Dirac points are shifted in energy in the {\em opposite} direction. As a result the edge modes acquire a dispersion,  Fig.\ \ref{fig1}e, which is now, crucially, the same for both edges. We thus obtain a system with two copropagating edge modes compensated by counter-propagating bulk modes.  Mathematically the requisite pseudoscalar potential follows from a term that is similar to the Haldane mass \cite{haldane}  for spinless fermions and describes a second-neighbor complex hopping between sites.  We show  that a variant of such a term is actually realized in transition metal dichalcogenide monolayers when one includes spin.

\begin{figure}[t]
\centering
\includegraphics[scale=0.35]{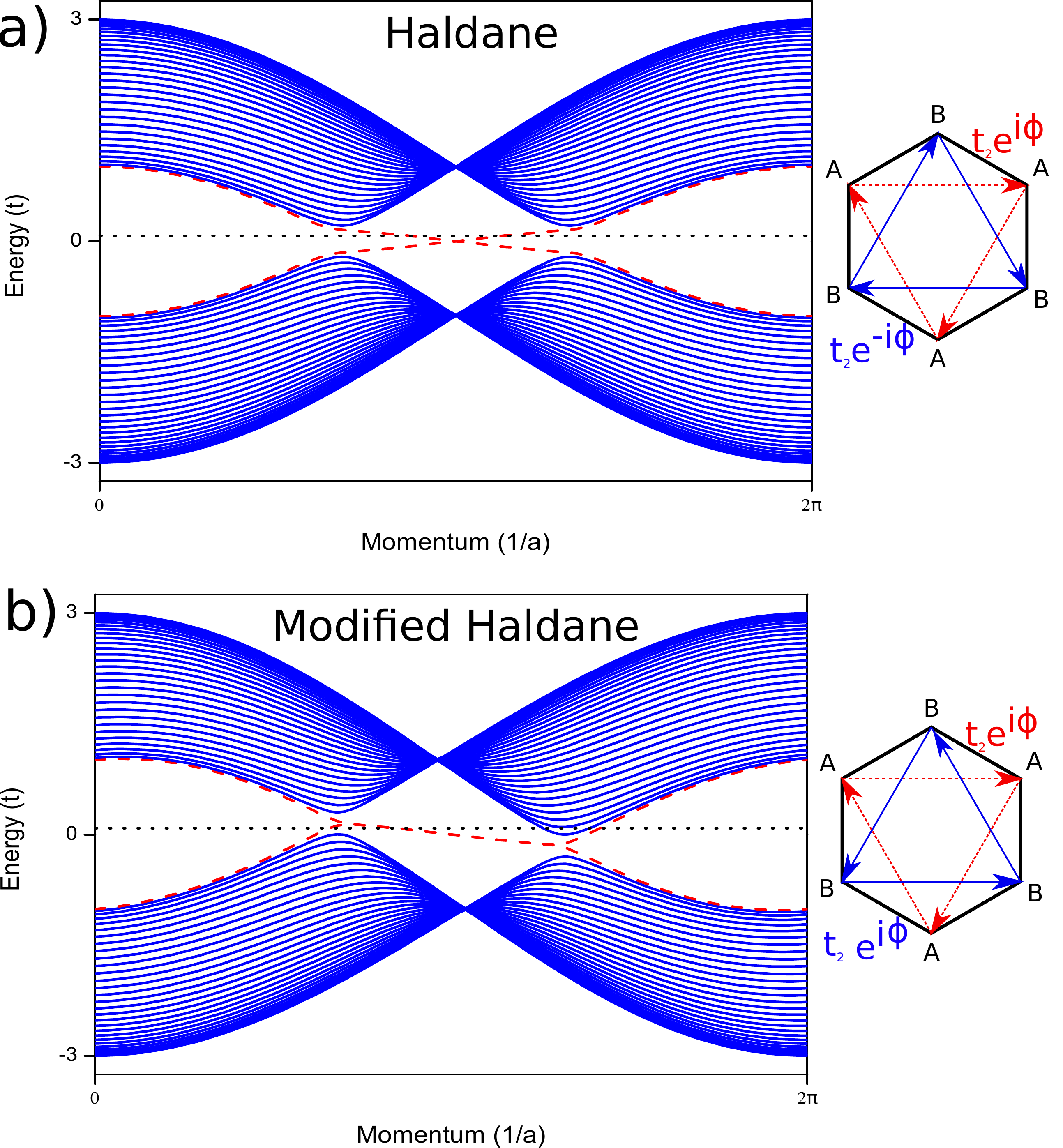}
\caption{Band structure of a zigzag nanoribbon of width $W=60$ (60 ZGNR) described by a) Haldane and b) modified Haldane Hamiltonian. $t_1$ is taken as the unit of energy, $t_2=0.03$ and $\phi=\pi/2$. Insets show the pattern of phases for the nnn hoping terms.}
\label{bands}
\end{figure}

\emph{Modified Haldane model.}-- We  seek a term to add to the graphene Hamiltonian which breaks time reversal symmetry ($\mathcal{T}$) and acts as a scalar potential with an opposite sign in each valley. In 1988 Haldane \cite{haldane} introduced a model realizing quantized Hall conductance without an external magnetic field, defined by the Hamiltonian
\begin{equation}
H = t_1\sum_{\langle i,j\rangle}c_i^\dag c_j + t_2\sum_{\langle \langle i,j\rangle\rangle}e^{-iv_{ij}\phi}c_i^\dag c_j,
\label{TBHM}
\end{equation}
where $c_i$ is an annihilation operator for  spinless fermion on site ${\bR}_i$ of the honeycomb lattice. The next nearest neighbor (nnn)  hopping breaks the $\mathcal{T}$ symmetry due to the phase $\phi$, and is different when it links two A atoms or B atoms ($v_{ij}=\pm 1$, see also inset \fref{bands}a). In the continuum theory, the Hamiltonian can be rewritten as
\begin{equation}
H = \hbar v_F(\sigma_x \tau_z q_x + \sigma_y \tau_0 q_y) +  t_2^a\sigma_z\tau_z + t_2^b\sigma_0\tau_0,
\label{HamHM}
\end{equation}
where $v_F$ is the Fermi velocity, $\bsig$ and $\btau$ are the Pauli matrices acting in the sublattice and valley spaces respectively, and $\bq$ is momentum relative to the Dirac points. In addition  $t_2^a=-3\sqrt{3}t_2\sin{\phi}$, and $t_2^b=-3t_2\cos{\phi}$.  In \fref{bands}a we display the band structure of the Haldane nanoribbon. When the Fermi energy $E_F$ lies close to zero energy it crosses the two edge states which connect the two Dirac points. One edge mode is right moving and the other left moving as expected of the chiral edge modes.

Importantly for our goal of constructing a pseudoscalar perturbation, we observe  in \eref{HamHM}  that the Haldane term $t_2^a\sigma_z\tau_z$ anticommutes with the rest of the Hamiltonian and acts therefore as a mass term. If we were to modify this term to act equally in both sublattices (i.e.\ replacing $\sigma_z \rightarrow \sigma_0$), then it would commute with the rest of the Hamiltonian and play the role of a pseudoscalar potential instead. On the lattice this can be achieved if  we make the nnn hopping term to act equally in both A and B sublattices, i.e. in \eref{TBHM} we set $v_{ij}=+ 1$ for all sites (see inset \fref{bands}a). With this change, the ``modified Haldane Model'' (mHM) at low energies becomes
\begin{equation}
H = \hbar v_F(\sigma_x \tau_z q_x + \sigma_y \tau_0 q_y) +  t_2^a\sigma_0\tau_z + t_2^b\sigma_0\tau_0.
\label{HamHMm}
\end{equation}
The excitation spectrum reads
\begin{equation}
E_\bq=\pm \hbar v_F\sqrt{q_x^2+q_y^2}+t_2^a\tau_z+t_2^b,
\label{ExHMm}
\end{equation}
showing that the two Dirac points are indeed offset in energy by $\pm t_2^a$ as outlined in Fig.\ \ref{fig1}e. On the basis of arguments presented in the introduction we  expect that mHM should exhibit antichiral edge modes.

In \fref{bands}b, we display the band structure of the mHM. The Dirac points are shifted in energy and the edge modes acquire dispersion with the same velocity, i.e.\ both edge modes propagate in the same direction. We see also that the Fermi energy crosses bulk modes as it should because  the total number of left- and right-moving modes must be the same. The important point is that these modes belong to the bulk and are therefore spatially separated from the edge modes. We therefore expect backscattering of the edge modes to be strongly suppressed. An electron in the edge that suffers a collision with an impurity cannot backscatter unless it moves to the bulk.

\emph{Results.}-- { In pristine graphene the zigzag edge zero modes are protected by the chiral symmetry ${\cal C}$: $\sigma_z H_0 \sigma_z = -H_0$ which allows a topological winding number to be defined \cite{ryu1,delplace1}. We review this topological protection in Supplementary Material \cite{supplement} and show that it applies to the modified Haldane model as well. In essence, because the mHM term is proportional to the unit matrix in the sublattice space, it does not modify the spinor structure of the wavefunctions compared to the pristine case (although it does change their energies). Since the topology is encoded in the wavefunctions we expect the edge modes persist, except that they  may now occur away from zero energy. This indeed is seen in \fref{bands}b.}

\begin{figure}[t]
\centering
\includegraphics[scale=0.35]{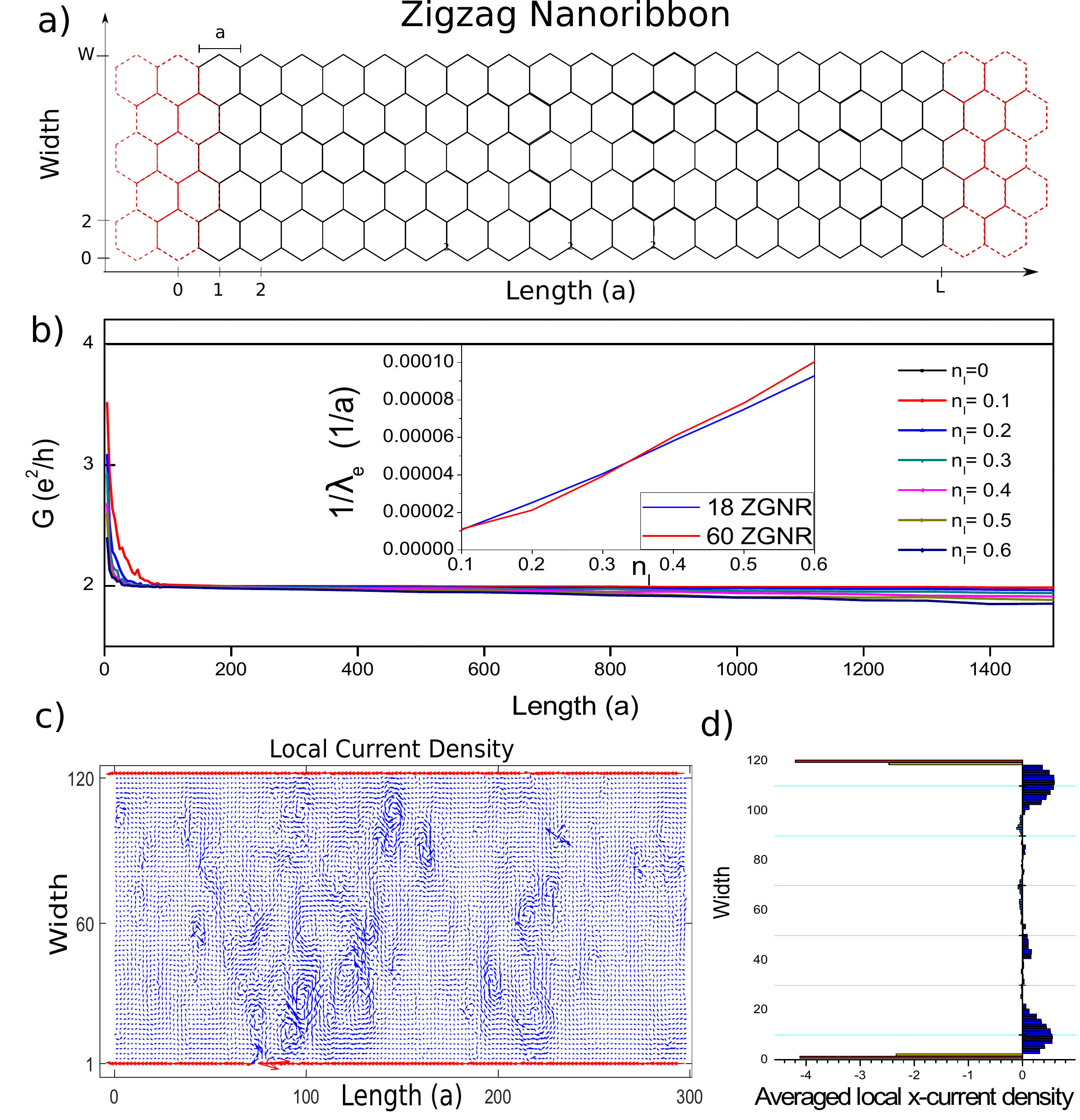}
\caption{a) Zigzag nanoribbon structure used in numerical simulations. b) Conductance as a function of $L$ for $W=60$,  $U=2.5$ and $E_F$  set close to zero. Different disorder concentrations are depicted. Inset: Inverse edge state localization length $\lambda_e$ estimated from the decay in conductance $G(L)\simeq 2(1-L/\lambda_e)$ in the regime $\lambda_b\ll L\ll\lambda_e$. We checked that the conductance is the same when the current flows from left to right and vice versa, as required by conductance reciprocity. c) Local current density (LCD) of a 120 ZGNR and $L$=300 for $E_F$ close to zero when $n_I=0.1$. d) Average of the LCD in the x-direction of c) as a function of the width of the ribbon.}
\label{c-l}
\end{figure}

To demonstrate the extreme robustness of the antichiral edge modes in mHM against disorder, we compute \cite{kwant1,kwant2} the conductance $G$ as a function of the length $L$ of the ribbon for different impurity concentrations $n_I$. We consider a system depicted in \fref{c-l}a, where the black part corresponds to the active region with disorder and the red part represents the contacts. Impurities are introduced as on-site potentials, whose energy value is randomly chosen in the interval $[-U,U]$. Simulations are repeated for different impurity configurations and the conductance is averaged over these. Impurity concentration $n_I$ is expressed as the number of defected sites divided by the total number of atoms in the system.

In \fref{c-l}b we plot conductance as a function of the length of the ribbon. For a clean sample ($n_I=0$) conductance (in units of $e^2/h$) equals to 4, independently of the length of the system. This can be explained by noting that for the adopted parameters  $E_F$ crosses two right-moving  bulk modes in addition to two right-moving  edge modes. When disorder is introduced, we observe an initial fast drop in conductance with $L$. We interpret this as Anderson localization \cite{AL1,AL2} of the bulk modes. However, contrary to the ordinary zigzag nanoribbon where conductance drops to zero \cite{length}, in the mHM conductance reaches the value of 2 and then remains essentially constant even for very long ribbons and very high values of disorder. This occurs because two edge modes (and two counter propagating bulk modes) remain delocalized and continue exhibiting ballistic transport. We emphasize that these modes are extremely robust. For instance, for  $n_I=0.6$ corresponding to very strong disorder with 60\% of sites containing defects (far higher than disorder levels in realistic graphene samples \cite{guinea}) and $L=1500$, conductance only decreases by about $5\%$.

These results suggest that mHM can be characterized by two localization lengths, one for the bulk modes which we define as $\lambda_{b}^{-1}=d\ln(G-2)/dL |_{L<\lambda_b}$ and one for the edge modes, $\lambda_{e}^{-1}=d\ln(G)/dL |_{L>\lambda_b}$. From \fref{c-l}b, we can estimate both localization lengths. We find that values obtained for $\lambda_b$ are similar to localization length results found in the literature for similar systems \cite{length}, while $\lambda_e$ is much longer, expressing the fact that the edge modes are extremely difficult to localize.

To elucidate the anomalously long localization length of the edge states we performed some analytical calculations. We expect these edge states to exponentially decay into the bulk, with the wave function of the form $\psi_{e}(\br)\simeq e^{ikx-y/\xi}/\sqrt{\xi L}$. If we assume that the counter-propagating bulk states are roughly constant throughout the sample with $\psi_{b}(\br)\simeq e^{-ikx}/\sqrt{W L}$, it is easy to  obtain an expression for the elastic scattering rate $\hbar/\tau = n_I (\xi a^4 u_0^2/2v\hbar W^2)$.
Here $a$ is the lattice constant, $u_0$ is the average impurity potential strength, $v$ is the velocity,  and $\xi$  the typical decay length for the edge states.
The localization length in one dimension is $\lambda_{e} = \pi l $, where $l=v\tau$  is the mean free path. We obtain
\begin{eqnarray}\label{loc1}
\lambda_e = \frac{2v^2\hbar^2 W^2 \pi}{ n_I \xi a^4 u_0^2}.
\end{eqnarray}
Inset to \fref{c-l}b confirms the expected dependence of $\lambda_e$ on the impurity concentration $n_I$.  The expected dependence on width $W$ is however not borne out; we find that the last remaining counter-propagating bulk states are not fully extended over the width of the sample but are instead concentrated along the edges with a characteristic lengthscale $\xi'\gg\xi$. This is illustrated in \fref{c-l}c and \fref{c-l}d. The expression Eq.\ (\ref{loc1}) thus holds if we replace $W\to\xi'$, producing a long localization length, which however does not diverge in the limit of a wide strip.  
{ At present we do not fully understand the origin of the lengthscale $\xi'$ but we note that it may herald interesting new physics in systems with antichiral edge modes to be explored in future studies.} 

\emph{Proposed experimental realizations.}-- Direct experimental realization of our model faces the same challenges as the Haldane model, which has only recently been realized using ultracold atoms in an optical lattice \cite{cold}. The same method could be used to realize the modified Haldane model. We also note a recent proposal invoking an iron-based ferromagnetic insulator lattice \cite{propose}. 

\begin{figure}[t]
\centering
\includegraphics[scale=0.3]{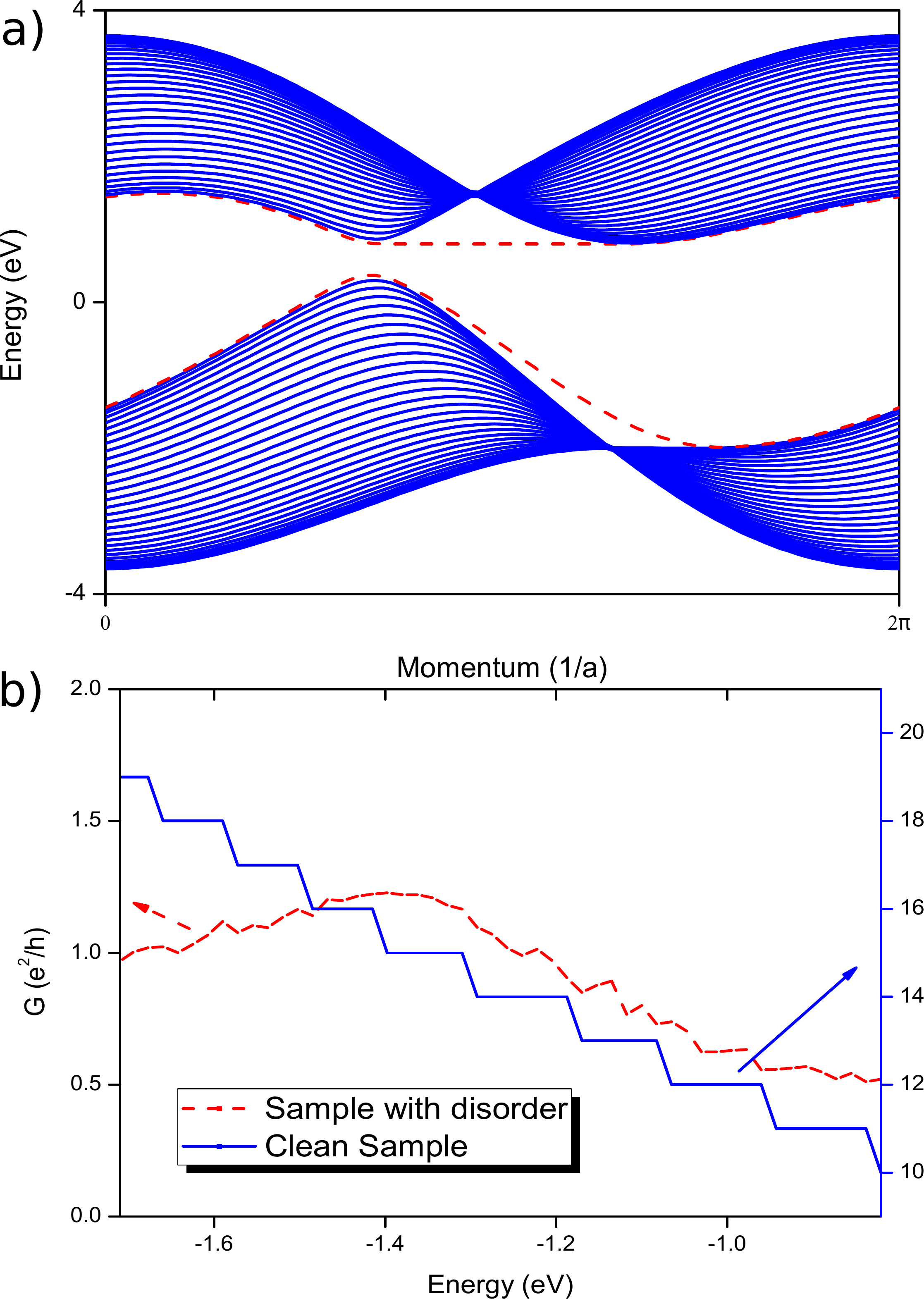}
\caption{a) WSe$_2$ band structure for a 60 ZGNR and $L=300$. Parameters used for the simulation are $t=1.19$\:eV, $m_S=0.8$\:eV and $\lambda=0.23$\:eV. b) Conductance for the interval of energies where the edge state exists in the valence band. The blue solid line corresponds to a clean sample while the red dashed line to a system with $n_I=0.1$.}
\label{MoS2}
\end{figure}

Another  route is based on the idea advanced by Kane and Mele \cite{kane} 
who noted that nnn tunneling amplitudes of the type required by the Haldane model can be supplied by spin-orbit coupling (SOC), creating effectively  two copies of the HM for two projections of the electron spin, conjugate under ${\cal T}$.  SOC is intrinsically very weak in graphene but the QSHE was  experimentally realized in HgTe quantum wells \cite{SQHEexp}.
Our strategy, therefore, is to generalize the model to spinfull fermions with SOC and obtain two copies of mHM conjugate under ${\cal T}$. In contrast to the HM, our model intrinsically breaks the inversion symmetry,  therefore we have to include SOC in a non centrosymmetric system. For that reason, hexagonal transition metal dichalcogenide  (TMD) monolayers $MX_2$ are excellent candidates. In their monolayer form, the $M=$ W, Mo atom is sandwiched between two $X=$ S, Se, Te atoms with $D^1_{3h}$ crystal structure \cite{structure}. TMDs have a similar band structure to graphene, with two nonequivalent Dirac points in the corners of the Brillouin zone but with a band gap due to the hybridization of the $d$ orbitals \cite{Mattheiss} and with stronger SOC since they are composed of heavy elements \cite{SOMS2}. Xiao {\em et al.} \cite{prl2012} proposed a low energy Hamiltonian for these materials and predicted selection rules for optical interband transitions, which have been tested experimentally \cite{exp1,exp2,exp3}. The Hamiltonian is
\begin{equation}
H = \hbar v_F(\sigma_x \tau_z q_x + \sigma_y \tau_0 q_y) -\lambda \tau_z \frac{\sigma_z-\sigma_0}{2} s_z+ \sigma_z m_S,
\label{HamPRL}
\end{equation}
where  $\lambda$ is the SOC parameter and $s_z$ represents the spin. We observe that in each spin sector SOC produces the desired pseudoscalar term proportional to $\tau_z\sigma_0$ in addition to the Haldane term $\tau_z\sigma_z$ and the inversion symmetry breaking ``Semenoff mass'' $m_S$. It is easy to construct the lattice version of \eref{HamPRL} and compute the band structure.  For the spin-up electrons this is displayed in \fref{MoS2}a  (the spin-down band structure is the same but reversed in momentum around the origin of the BZ). For the simulation, we chose $M=$ W and $X=$ Se since WSe$_2$ has the strongest SOC of all TMD monolayers \cite{data}. 

There are some obvious differences between \fref{MoS2}a and the ideal mHM band structure \fref{bands}b. Most importantly WSe$_2$ exhibits full bulk bandgap whereas mHM remains gapless. { Nevertheless WSe$_2$ shows edge modes that may be regarded as descendants of  those in mHM. Indeed it is easy to see that one can evolve the band structure in \fref{bands}b into \fref{MoS2}a by gradually turning on the Haldane and Semenoff mass parameters.  This illustrates the robustness of the topological protection even with ${\cal C}$ strongly broken.} If the Fermi energy crosses the WSe$_2$  edge mode in the valence band, similarly to our model, a current will flow along one edge of the sample with the countercurrent returning through the bulk. In this sense the WSe$_2$ zigzag nanoribbon realizes the physics of the antichiral edge modes and we expect them to be robust against disorder. \fref{MoS2}b shows the conductance as a function of $E_F$ for WSe$_2$. We find that disorder quickly localizes  all the bulk modes and only the edge mode survives, leading to conductance close to $e^2/h$. Interestingly, the edge currents in WSe$_2$ are spin polarized which could be useful in spintronic applications.

\emph{Conclusions.}-- We discussed 2D systems where nearly dissipationless currents occur because left and right moving modes are segregated to the edges and the bulk of the sample respectively. For that reason, backscattering is suppressed, even for samples with high impurity concentrations. A simple system showing this behavior can be constructed by slightly modifying the well known Haldane model \cite{haldane}. A variant  of this model (for spinfull fermions) is approximately realized in TMD monolayers and can be used to experimentally test the physics of antichiral edge modes.

\begin{acknowledgments}
\emph{Acknowledgments.}-- We thank NSERC, CIfAR, and Max Planck–-UBC Centre for Quantum Materials for support. E. C. acknowledges funding from Fondo Europeo de Desarrollo Regional (FEDER), the ``Ministerio de Ciencia e Innovaci\'{o}n'' through the Spanish Project TEC2015-67462-C2-1-R, the Generalitat de Catalunya (2014 SGR-384), and from the EU’s Horizon 2020 research program under grant agreement no.\ 696656.
\end{acknowledgments}

\newpage

\section*{SUPPLEMENTAL MATERIAL}

\section*{Topological protection of the edge modes}
To clarify the topological protection of the zigzag edge modes in graphene it is useful to write the lattice Hamiltonian Eq. (1) of the main text in the momentum space, $H=\sum_\bk\psi_\bk^\dagger h_\bk\psi_\bk$. Here $\psi_\bk=(a_\bk,b_\bk)^T$ and $a_\bk$, $(b_\bk)$ annihilate electrons in sublattice A (B) of the graphene honeycomb lattice. The Bloch Hamiltonian reads
\begin{equation}
\label{s1}
h_\bk=\begin{pmatrix}
\lambda_\bk  & \tau_\bk \\
\tau_\bk^* & \lambda_\bk  \end{pmatrix}, \ \ \ \ 
\end{equation}
where $\tau_\bk=t_1(1+e^{i\bk\cdot \ba_1}+e^{i\bk\cdot \ba_2})$ describes the pristine graphene and 
$\lambda_\bk=t_2[\sin{\bk\cdot\ba_1}-\sin{\bk\cdot\ba_2}-\sin{\bk\cdot(\ba_1-\ba_2)}]$  is the mHM term.  Here $\ba_p$ denote the primitive lattice vectors $\ba_1=a(1,0)$ and $\ba_2=a(1/2,\sqrt{3}/2)$ with $a$ the lattice constant.

\begin{figure}[t]
\includegraphics[scale=0.35]{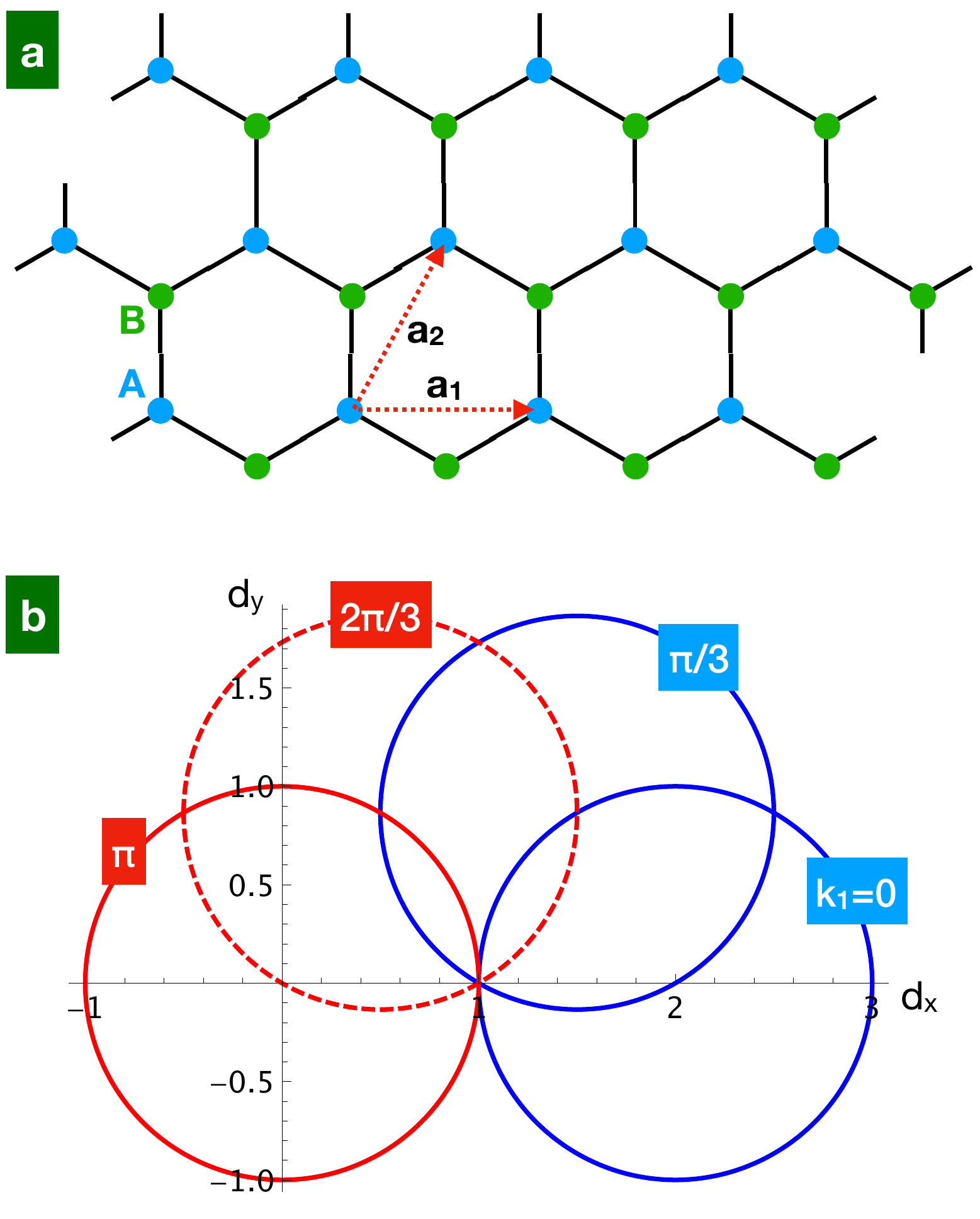}
\caption{a) Graphene lattice structure with zigzag edge along $\ba_1$
  direction. b) Parametric plot of the components of the Hamiltonian
  vector $\bd_{k_1}(k_2)$ as a function of $k_2$ for fixed values of
  $k_1$. We observe that for $|k_1|>2\pi/3$ the vector encircles the origin
  once, indicating a topological phase with winding number $\nu=1$.
 For  $|k_1|<2\pi/3$ the
  winding number is 0, indicating a trivial phase. Dashed line for
  $k_1=2\pi/3$ marks the critical point where vector $\bd$ passes
  through the origin, indicating a gapless spectrum.}
\label{fig_supp}
\end{figure}

We first discuss pristine graphene ($t_2=\lambda_\bk=0$) in a geometry
of a ribbon  with zigzag edges, infinite  along the $\ba_1$-direction
as depicted in Fig. \ref{fig_supp}a. Much of what we need is already
worked out in the existing literature
\cite{ryu1,delplace1,kharitonov1} and here we only give a brief
review. To understand the topological origin of the edge modes it is
instructive to consider first periodic boundary conditions along the
$\ba_2$-direction, i.e. an infinitely long cylinder. For each fixed
$k_1=\bk\cdot\ba_1$ in the boundary Brillouin zone we can view $h_\bk$
as describing a 1D crystal along the $\ba_2$-direction. We define this
system by a 1D Bloch Hamiltonian $h_{k_1}(k_2)\equiv
h_{(k_1,k_2)}$. Except when $k_1$ coincides with one of the graphene's
nodal points this Hamiltonian represents a gapped 1D insulator. The Hamiltonian respects the chiral symmetry 
\begin{equation}
\label{s2}
{\cal C}: \ \sigma_z  h_{k_1}(k_2) \sigma_z = -h_{k_1}(k_2).
\end{equation}
The chiral symmetry defines a symmetry class AIII which is known to
have integer topological classification in 1D \cite{altland1997}. The
relevant topological invariant is the winding number \cite{zak1989}, 
\begin{equation}
\label{s3}
\nu_{k_1}=i\oint_{\rm BZ}{dk_2\over \pi}\langle u_{k_1}(k_2)|\partial_{k_2}u_{k_1}(k_2)\rangle,
\end{equation}
where $|u_{k_1}(k_2)\rangle$ is an eigenstate of $h_{k_1}(k_2)$. The
corresponding physical quantity is the polarization $P={1\over 2}e\nu$
\cite{Kane_rev} which in 1D corresponds to the end charge $Q_{\rm end}$. Therefore, nonzero index $\nu$ implies electrical charge $ Q_{\rm end}=\pm {1\over 2}e\nu$ bound to each end of the 1D system.

For a generic gapped  $2\times 2$ matrix Hamiltonian $h(k)=\bsig\cdot\bd(k)$ the
winding number (\ref{s3}) is equal to $\Omega/2\pi$ where $\Omega$ is
the area on a unit sphere swept by a unit vector $\hat{\bd}(k)
=\bd(k)/|\bd(k)|$ as $k$ traverses the BZ. For a system with chiral
symmetry ${\cal C}$ vector $\bd(k)$ necessarily lies in the $x$-$y$
plane and $\hat{\bd}(k)$ is confined to the equator of the unit
sphere.  Then, index $\nu$ simply counts the number of times $\bd(k)$
winds around the origin and is therefore constrained to be an integer. 1D
systems with the chiral symmetry  ${\cal C}$ thus exhibit
quantized polarization $P$ in units of $e/2$. As it is well known from
the study of the Su-Schrieffer-Heeger (SSH) model \cite{SSH}, fractional values of the polarization are associated with zero modes localized near the end of the system. It is these SSH zero modes that furnish connection with the zigzag edges of graphene.

It is a simple matter to calculate the winding number of the 1D system defined by $h_{k_1}(k_2)$ for each fixed $k_1$ in the boundary BZ $(-\pi/a,\pi/a)$ using Eq. (\ref{s3}). One finds
\begin{equation}
\label{s4}
\nu_{k_1}=\biggl\{\begin{array}{cc}
0, & |k_1|< K_1, \\
1, & |k_1|>K_1,
\end{array}
\end{equation}
where $K_1={2\pi\over 3a}$ is the position of the Dirac point projected onto the boundary BZ. Alternately, one can consider the behavior of vector $\bd_{k_1}(k_2)$ as illustrated in Fig. \ref{fig_supp}b.  These considerations imply that closing of the gap at the Dirac point may be seen as marking a topological phase transition where the index $\nu_{k_1}$ changes its value from $0$ to $1$. In accord with our previous discussion we therefore expect topologically protected zero modes to appear at the zigzag edge of graphene for $|k_1|>K_1$. This is indeed confirmed by direct numerical calculations for graphene in the strip geometry \cite{ryu1,delplace1,kharitonov1}.

When $t_2\neq 0$ the Hamiltonian breaks the chiral symmetry $\cal C$
due to the nonzero $\lambda_\bk$ in Eq.0(\ref{s1}) . It is important
to note that the symmetry breaking term is proportional to the unit
matrix in the sublattice space. Therefore while the  eigenenergies are
shifted the Bloch eigenstates  $|u_{k_1}(k_2)\rangle$ remain
unchanged. Because the winding number in Eq. (\ref{s3}) depends only
on the eigenstates it too must remain the same. We conclude that our
modified Haldane model has exactly the same topological structure in
relation to the zigzag edge states as pristine graphene.  

We thus expect protected edge modes to exist in mHM in the same range
of $k_1$ between the projected Dirac points. 
Because the mHM term breaks $\cal C$ the edge modes no longer occur
at exactly zero energy but as apparent in Fig.\ 2b of the main text
they remain robustly present, connecting between the Dirac points.
 Residual symmetries of the system
furthermore guarantee that $h_{k_1}(k_2)$ retains the chiral symmetry
at $k_1=0$ and $k_1=\pi/a$. At these points fractional polarization implies exact zero modes and indeed we see that the edge state crosses zero energy exactly midway between the two Dirac points in Fig. 2b.

To summarize, edge modes along the zigzag edges of graphene are topologically protected by winding numbers associated with the family of 1D Hamiltonians defined for fixed crystal momenta parallel to the edge. This protection extends to graphene with the modified Haldane term which has been designed to offset its two Dirac points in energy and make the edge modes dispersive.  


\begin{thebibliography}{99}
\bibitem{remark1} For helical edge states unbroken time reversal symmetry and negligible interaction strength must be maintained to achieve  dissipationless transport.
\bibitem{kwant1}
Numerical simulations were performed with the Kwant tight-binding code.
\end{thebibliography}
\end{document}